\documentclass[12pt]{article}
\usepackage{epsfig}
\usepackage{slashed}
\def\br(#1,#2){\left\langle#1#2\right\rangle}
\def\sq(#1,#2){\left[#1#2\right]}
\def\s(#1,#2){s_{#1 #2}}
\def\t(#1,#2,#3){s_{#1 #2 #3}}

\begin{document}
\begin{titlepage}

\hspace*{\fill}\parbox[t]{5cm}
{hep-ph/0410370 \\
\\
\today} \vskip2cm
\begin{center}
{\Large \bf Symmetries of the Standard Model} \\
\medskip
\bigskip\bigskip\bigskip\bigskip
{\large {\bf Scott Willenbrock}} \\
\bigskip\bigskip\medskip
Department of Physics, University of Illinois at Urbana-Champaign \\
1110 West Green Street, Urbana, IL\ \ 61801 \\ \bigskip
\end{center}

\bigskip\bigskip\bigskip

\begin{abstract}
I present an overview of the standard model, concentrating on its
global continuous symmetries, both exact and approximate.  There
are four lectures, dedicated to spacetime symmetry, flavor
symmetry, custodial symmetry, and scale symmetry.  Topics include
Weyl, Majorana, and Dirac spinors; massive neutrinos; electroweak
symmetry breaking; effective field theory; and the hierarchy
problem.
\end{abstract}

\end{titlepage}

I was asked to give four lectures entitled {\it Introduction to
the Standard Model}.  However, I suspect all TASI students already
have some familiarity with the standard model, so I do not intend
to start from scratch.  Instead, I would like to talk about some
advanced topics within the standard model, ones that are not
necessarily covered in a first course. After I finished preparing
these lectures, I noticed that there was an underlying theme of
symmetry, hence the title of these lectures.

The first two lectures are aimed at understanding neutrinos.
Neutrino mass is definitely physics beyond the standard model.  My
aim is to convince you of this.  The first lecture is a general
introduction to fermion fields, in particular their transformation
properties under spacetime symmetry. The second lecture is
concerned with flavor symmetry in both the quark and lepton
sector.

The last two lectures are about the Higgs sector of the standard
model.  Lecture 3 discusses custodial symmetry both within the
standard model and more generally. Lecture 4 discusses why the
standard Higgs model is problematic when you embed it in a more
fundamental theory characterized by a very high energy scale, such
as in grand unification.

Although the history of particle physics is fascinating, my
presentation will not follow the historical development of these
topics.  Instead, I will try to present things in a logical
manner, deriving as many things as I can from scratch.  Some of
the details I leave to you in the form of exercises; solutions are
included in an Appendix.  Occasionally I will make a leap, but I
will try to point you to places where you can fill in the gap.

\section{Weyl, Majorana, Dirac}
\label{sec:weyl}

We all hear (and use) the names Weyl, Majorana, and Dirac in the
context of fermions, but there tends to be some confusion about
them.  In this lecture I hope to clarify their use.  I will
discuss the Lorentz-transformation properties of fermions, but I
will stop short of actually quantizing the fields in terms of
creation and annihilation operators.  Thus I will only be dealing
with the classical aspects of these fields.  The quantized fermion
field is dealt with in many textbooks.  I will usually work with
the notation and conventions of Peskin and Schroeder
\cite{Peskin:1995ev}, which is an excellent introduction to
quantum field theory.

The generators of the Lorentz group are the rotations, $J_i$, and
the boosts, $K_i$.  They satisfy the algebra
\cite{Weinberg:1995mt}
\begin{eqnarray*}
&&[J_i,J_j]=i\epsilon_{ijk} J_k \\
&&[K_i,K_j]=-i\epsilon_{ijk} J_k \\
&&[J_i,K_j]=i\epsilon_{ijk} K_k\;.
\end{eqnarray*}
The $J_i$ are Hermitian, and the $K_i$ are anti-Hermitian. The
$J_i$ satisfy the algebra of the rotation group, $SU(2)$.  The
last commutation relation expresses the fact that a boost
transforms as a three-vector under rotations. To disentangle the
algebra, define the Hermitian generators
\begin{eqnarray*}
A_i&=&{1\over 2} (J_i+iK_i)\\
B_i&=&{1\over 2} (J_i-iK_i)\;.
\end{eqnarray*}
\\[7pt]
{\em Exercise 1.1 - Show that the $A_i$ and $B_i$ satisfy the
algebra}
\begin{eqnarray*}
&&[A_i,A_j] = i\epsilon_{ijk} A_k \\
&&[B_i,B_j] = i\epsilon_{ijk} B_k \\
&&[A_i,B_j ]= 0\;.
\end{eqnarray*}
The algebra for the $A_i$ and the $B_i$ is that of $SU(2)$, and
the two algebras are independent.  We have thus shown that the
Lorentz group, $SO(3,1)$, is locally isomorphic to $SU(2)\times
SU(2)$.

Representations of $SU(2)$ are familiar from the representations
of the rotation group: each representation is label by ``spin,''
which can have integer or half-integer values.  Thus the
representations of the Lorentz group are labeled $(a,b)$, where
$a,b=1/2, 1, 3/2,...$.  The simplest representation is $(0,0)$,
which corresponds to a scalar field.  The simplest nontrivial
representation is $(1/2,0)$, which corresponds to a Weyl spinor,
$\chi$.  The generators are
\begin{eqnarray*}
A_i&=&{1\over 2}\sigma_i \\
B_i&=&0
\end{eqnarray*}
which corresponds to
\begin{eqnarray*}
J_i&=&{1\over 2}\sigma_i \\
iK_i&=&{1\over 2}\sigma_i\;.
\end{eqnarray*}
Hence a Weyl spinor is a 2-component object that transforms under
rotations and boosts as
\begin{eqnarray}
\chi &\to &e^{-{i\over 2} \sigma\cdot\theta}\chi \nonumber\\
\chi &\to &e^{-{1\over 2} \sigma\cdot\eta}\chi\label{eq:rotboost}
\end{eqnarray}
where $\eta$ is the rapidity, which is related to the velocity by
$\beta=\tanh\eta$.  Here I am taking the ``active'' viewpoint,
transforming the spinor while keeping the coordinate system fixed.
The transformation under rotations shows that a Weyl spinor
carries spin 1/2.
\\[7pt]
{\em Exercise 1.2 - Calculate the transformations under rotations
and boosts of the $(0,1/2)$ representation.}
\\[7pt]
\indent Let us construct a Lorentz-invariant mass term from a
single Weyl spinor.  This is called a Majorana mass.  It is given
by
\begin{equation}
{\cal L}={1\over 2}m(\chi^T\epsilon\chi + h.c.)
\label{eq:Maj}
\end{equation}
where $\epsilon \equiv i\sigma_2$ is
the $2\times 2$ antisymmetric matrix.  Let's show that this mass
term is Lorentz invariant. Denote a Lorentz transformation acting
on $\chi$ by the matrix $M$, where
\begin{equation}
M=e^{-{i\over 2} \sigma\cdot\theta}\;\; {\rm or}\;\; e^{-{1\over
2} \sigma\cdot\eta}\;.
\end{equation}
Thus under a Lorentz transformation,
\begin{equation}
\chi^T\epsilon\chi\to\chi^TM^T\epsilon M\chi\;.
\label{eq:Lorentz}
\end{equation}
Displaying indices,
\begin{equation}
(M^T)_{\alpha\beta}\epsilon_{\beta\gamma}M_{\gamma\delta}=
\epsilon_{\beta\gamma}M_{\beta\alpha}M_{\gamma\delta}=\epsilon_{\alpha\delta}\det
M=\epsilon_{\alpha\delta}\label{eq:MTeM}
\end{equation}
where the last step uses the fact that $\det M=1$. Thus
$M^T\epsilon M = \epsilon$, which completes the proof in
Eq.~(\ref{eq:Lorentz}) that $\chi^T\epsilon\chi$ is Lorentz
invariant.  Incidentally, this also shows that the Lorentz group
is locally isomorphic to $SL(2,\mathcal{C})$, the group of
$2\times 2$ complex matrices of unit determinant
\cite{Wess:1992cp}.

Although a Majorana mass is less familiar than a Dirac mass, it is
actually a more basic quantity, constructed from a single Weyl
spinor.  In this sense a Majorana mass is the simplest fermion
mass term.  However, if $\chi$ carries an unbroken global or local
$U(1)$ charge, a Majorana mass is forbidden, since it would
violate this symmetry.  Thus none of the fermions of the standard
model (except neutrinos) can have a Majorana mass, since they
carry electric charge. More generally, if $\chi$ transforms under
a complex or pseudoreal representation of an unbroken global or
local internal symmetry, a Majorana mass is forbidden.  Let
\begin{equation}
\chi\to U\chi
\end{equation}
where $U$ is a unitary transformation acting on a set of Weyl
spinors. The mass term in Eq.~(\ref{eq:Lorentz}) transforms as
\begin{equation}
\chi^T\epsilon\chi\to\chi^TU^T\epsilon U\chi=\chi^T\epsilon
U^TU\chi
\end{equation}
where in the last step I have used the fact that $U$ and
$\epsilon$ act on different spaces.  This is invariant only if
$U^TU=1$, which is true only if the unitary transformation $U$ is
real ($U^*=U$) \cite{Peskin:1995ev}.

Physically, a fermion with a Majorana mass is its own antiparticle
\cite{Kayser:1989iu}.  It is referred to as a Majorana fermion. It
cannot carry an unbroken global or local $U(1)$ charge (or, more
generally, transform under a complex or pseudoreal representation)
because a particle and an antiparticle must carry opposite charge.
\\[7pt]
{\em Exercise 1.3 - Gluinos are hypothetical Majorana fermions
that are the superpartners of gluons.  Why can they carry color
charge?}
\\[7pt]
{\em Exercise 1.4 - Consider a Weyl fermion that transforms under
the defining representation of an unbroken $SU(2)$ group.  Show
that the Majorana mass term}
\begin{equation}
{\cal L}={1\over 2}m(\epsilon_{ab}\chi^{aT}\epsilon\chi^{b} +
h.c.)
\end{equation}
{\em is invariant under the $SU(2)$ symmetry.  However, show that
this term vanishes.}
\\[7pt]
\indent If a Weyl fermion transforms under a complex or pseudoreal
representation of an unbroken global or local symmetry, then we
need to introduce a second Weyl fermion that transforms under the
complex-conjugate representation in order to construct a mass
term.  This is a Dirac mass.  Let $\chi,\xi$ transform under the
$(1/2,0)$ representation of the Lorentz group, and transform under
some unbroken global or local symmetry as
\begin{eqnarray*}
\chi&\to &U\chi \\
\xi&\to &U^*\xi\;.
\end{eqnarray*}
Then a Lorentz-invariant mass term may be formed which respects
the unbroken symmetry,
\begin{equation}
{\cal L}=m(\xi^T\epsilon\chi + h.c.)
\label{eq:Dirac}
\end{equation}
since
\begin{equation}
\xi^T\epsilon\chi \to \xi^TU^\dagger\epsilon U\chi =
\xi^T\epsilon\chi\;.
\end{equation}
Thus it takes two Weyl spinors to construct a Dirac mass.  A
fermion with a Dirac mass is called a Dirac fermion.

That's pretty much all the physics in this lecture; the rest is
mathematics, in order to make contact with the standard way of
treating this subject.  However, there is a result at the end of
the lecture that will be of physical interest in Lecture 2.

Thus far we have constructed a Majorana mass from a Weyl spinor,
and a Dirac mass from a pair of Weyl spinors. Let us now introduce
a new object, the Dirac spinor, which is a four-component object
constructed from a pair of $(1/2,0)$ Weyl spinors $\chi,\xi$ via
\begin{equation}
\psi=\left(\begin{array}{c}\chi\\
\epsilon\xi^*\end{array}\right)\;.
\end{equation}
In terms of a Dirac spinor, a Dirac mass is written in the
familiar form
\begin{eqnarray*}
{\cal
L}=-m\bar\psi\psi&=&-m\left(\chi^\dagger,-\xi^T\epsilon\right)\left(\begin{array}{cc} 0 & 1 \\
1 & 0 \end{array}\right)\left(\begin{array}{c}\chi\\
\epsilon\xi^*\end{array}\right)\\
&=&m(\xi^T\epsilon\chi-\chi^\dagger\epsilon\xi^*)
\end{eqnarray*}
where the final expression is identical with Eq.~(\ref{eq:Dirac}).
We are using a specific basis for the Dirac gamma matrices, the
so-called Weyl or chiral basis,
\begin{equation}
\gamma^0=\left(\begin{array}{cc} 0 & 1 \\
1 & 0 \end{array}\right) \;\;\;
\gamma^i=\left(\begin{array}{cc} 0 & \sigma^i \\
-\sigma^i & 0 \end{array}\right) \;\;\;
\gamma_5=\left(\begin{array}{cc} -1 & 0 \\
0 & 1 \end{array}\right) \;\;\;,
\end{equation}
where each entry in the above matrices is itself a $2\times 2$
matrix.  In this basis, the chiral projection operator $(1\pm
\gamma_5)/2$ projects out the Weyl spinors,
\begin{equation}
\psi=\frac{1-\gamma_5}{2}\psi+\frac{1+\gamma_5}{2}=\psi_L+\psi_R
\end{equation}
where
\begin{eqnarray*}
\psi_L&=&\left(\begin{array}{c}\chi
\\0\end{array}\right)\\
\psi_R&=&\left(\begin{array}{c}0
\\ \epsilon\xi^*\end{array}\right)\;.
\end{eqnarray*}
Thus $\psi_L$ is the four-component Dirac-spinor version of the
Weyl spinor $\chi$, and similarly for $\psi_R$ and
$\epsilon\xi^*$.
\\[7pt]
{\em Exercise 1.5 - If $\xi$ transforms under the $(1/2,0)$
representation of the Lorentz group, show that $\epsilon\xi^*$
transforms under the $(0,1/2)$ representation.  [Hint: recall
$M^T\epsilon M=\epsilon$, which we derived in
Eq.~(\ref{eq:MTeM})].}
\\[7pt]
Thus a Dirac spinor transforms under the $(1/2,0)\oplus (0,1/2)$
representation of the Lorentz group, corresponding to
$\psi=\psi_L+\psi_R$.

While a Dirac spinor is composed of two Weyl spinors, a Majorana
spinor is a four-component object composed of a single Weyl
spinor,
\begin{equation}
\psi_M=\left(\begin{array}{c}\chi
\\ \epsilon\chi^*\end{array}\right)\;.
\label{eq:Majorana}
\end{equation}
Thus it is simply a four-component version of a Weyl spinor.
\\[7pt]
{\em Exercise 1.6 - Show that}
\begin{equation}
{\cal L}=-{1\over 2}m\bar\psi_M\psi_M\label{eq:LMajorana}
\end{equation}
{\em is a Majorana mass term.}
\\[7pt]
\indent We can find even more ways of writing fermion mass terms
by introducing the charge-conjugation matrix $C$, which in the
Weyl or chiral representation of the Dirac matrices is
\begin{equation}
C=\left(\begin{array}{cc} -\epsilon & 0 \\
0 & \epsilon \end{array}\right)\;.
\end{equation}
Given a Dirac spinor $\psi$, we can form the conjugate spinor via
\begin{eqnarray*}
\psi^c&\equiv &C\gamma^0\psi^* \\
&=&\left(\begin{array}{cc} -\epsilon & 0 \\
0 & \epsilon \end{array}\right)\left(\begin{array}{cc} 0 & 1 \\
1 & 0 \end{array}\right)\left(\begin{array}{c}\chi^*\\ \epsilon\xi\end{array}\right)\\
&=&\left(\begin{array}{c}\xi\\ \epsilon\chi^*\end{array}\right)\;.
\end{eqnarray*}
Thus
\begin{eqnarray*}
\psi_L&=&\left(\begin{array}{c}\chi\\ 0\end{array}\right)\\
 \\
(\psi^c)_L&=&\left(\begin{array}{c}\xi\\ 0\end{array}\right)
=(\psi_R)^c\;.
\end{eqnarray*}
Note that the last relation implies that conjugation and chiral
projection do not commute.
\\[7pt]
{\em Exercise 1.7 - Show that}
\begin{equation}
{\cal L}=-m((\psi^c)_L^TC\psi_L + h.c.)\label{eq:LDiracpsic}
\end{equation}
{\em is a Dirac mass.}
\\[7pt]
{\em Exercise 1.8 - Show that}
\begin{equation}
{\cal L}=-{1\over 2}m(\psi_L^TC\psi_L +
h.c.)\label{eq:LMajoranaDirac}
\end{equation}
{\em is a Majorana mass. Thus one can write a Majorana mass in
terms of a Dirac spinor.}
\\[7pt]
{\em Exercise 1.9 - Show that}
\begin{equation}
\psi_M^c=\psi_M\;.
\end{equation}
{\em This is called the Majorana condition.}
\\[7pt]
\indent In Exercise 1.8 we wrote a Majorana mass in terms of a
Dirac spinor.  Can we write a Dirac mass in terms of Majorana
spinors? Yes, and this is the physically relevant result I
promised you earlier in the lecture.  Consider a Dirac mass
written in terms of a Dirac spinor,
\begin{equation}
{\cal L}=-m\bar\psi\psi = -{1\over
2}m(\bar\psi\psi+\bar\psi^c\psi^c)
\label{eq:Dirac2}
\end{equation}
where I'll let you verify the last equality.  Now define the
Majorana spinors
\begin{eqnarray*}
\psi_M^1&\equiv &{1\over \sqrt 2}(\psi+\psi^c)\\
\psi_M^2&\equiv &{1\over \sqrt 2}(\psi-\psi^c)\;.
\end{eqnarray*}
Then Eq.~(\ref{eq:Dirac2}) can be written
\begin{equation}
{\cal L}=-{1\over
2}m(\bar\psi_M^1\psi_M^1+\bar\psi_M^2\psi_M^2)\;.
\end{equation}
Thus a Dirac fermion is equivalent to two degenerate Majorana
fermions. However
\begin{eqnarray*}
(\psi_M^{1})^c&=&{1\over \sqrt 2}(\psi^c+\psi)=\psi_M^1\\
(\psi_M^{2})^c&=&{1\over \sqrt 2}(\psi^c-\psi)=-\psi_M^2\;.
\end{eqnarray*}
Thus the two Majorana spinors have the opposite sign under charge
conjugation.  The Majorana spinor $\psi_M^2$ is therefore of the
form
\begin{equation}
\psi_M^2=\left(\begin{array}{c}\chi\\ -\epsilon\chi^*\end{array}\right)\\
\end{equation}
which is a generalization of the construction in
Eq.~(\ref{eq:Majorana}). We will see at the end of Lecture 2 the
physical significance of these results.

The various ways we have learned to write Majorana and Dirac
masses in terms of Weyl, Majorana, and Dirac spinors are collected
in Table~\ref{tab:mass}.  As a final exercise, I invite you to
derive the one entry in this Table that we have not yet discussed.
\\[7pt]
{\em Exercise 1.10 - Show that}
\begin{equation}
{\cal L}=-{1\over 2}m(\overline{(\psi^c)}_R\psi_L +
h.c.)\label{eq:LMajoranapsic}
\end{equation}
{\em is a Majorana mass.}
\\[7pt]
This form for the Majorana mass in terms of Dirac spinors is used
quite commonly \cite{Kayser:1989iu}.

\begin{table}[tbh]
\begin{center}
\begin{tabular}{|c||c|c|}
\hline
spinor & Majorana mass & Dirac mass \\
\hline Weyl & ${1\over 2}m(\chi^T\epsilon\chi + h.c.)$ & $m(\xi^T\epsilon\chi + h.c.)$ \\
Majorana & $-{1\over2}m\bar\psi_M\psi_M$
& $-{1\over 2}m(\bar\psi_M^1\psi_M^1+\bar\psi_M^2\psi_M^2)$ \\
Dirac & $-{1\over 2}m(\psi_L^TC\psi_L + h.c.)$ & $-m((\psi^c)_L^TC\psi_L + h.c.)$ \\
Dirac & $-{1\over 2}m(\overline{(\psi^c)}_R\psi_L + h.c.)$ & $-m\bar\psi\psi$ \\
\hline \hline
\end{tabular}
\caption{A Majorana mass and a Dirac mass may be constructed from
Weyl, Majorana, or Dirac spinors.}
\end{center}
\label{tab:mass}
\end{table}

\section{Flavor symmetry}
\label{sec:flavor}

In Table~\ref{tab:sm} I list the fermion fields that make up the
standard model, along with their $SU(3)\times SU(2)_L\times
U(1)_Y$ quantum numbers.  The index $i=1,2,3$ on each field refers
to the generation.  I list the fields in terms of left-chiral
Dirac spinors, which, as we saw in the previous section, is the
four-component version of a Weyl spinor.  For example, one should
think of
\begin{eqnarray*}
u_L&=&\left(\begin{array}{c}\chi\\ 0\end{array}\right)\\
(u^c)_L&=&\left(\begin{array}{c}\xi\\ 0\end{array}\right)
\end{eqnarray*}
where $\chi,\xi$ are $(1/2,0)$ Weyl spinors.  {\it A priori} $u_L$
and $(u^c)_L$ are totally independent, despite their names. They
are only named thusly because we know they will eventually pair up
to make a Dirac spinor
\begin{equation}
u=\left(\begin{array}{c}\chi\\
\epsilon\xi^*\end{array}\right)\;. \label{eq:Diracfermion}
\end{equation}
Working in terms of only left-chiral fields is particularly useful
for grand unification, where we attempt to combine fermions into
representations of a group that contains the standard model as a
subgroup.  Since left-chiral fields all transform the same way
under the Lorentz group, combining them respects Lorentz
invariance.

\begin{table}[tb]
\begin{center}\begin{tabular}[7]{cccccccc}
&&&&$\underline{SU(3)}$&$\underline{SU(2)_L}$&$\underline{U(1)_Y}$\\
\\
$Q_L^i=$&$\left(\begin{array}{l}u_L\\d_L\end{array}\right)$
&$\left(\begin{array}{l}c_L\\s_L\end{array}\right)$
&$\left(\begin{array}{l}t_L\\b_L\end{array}\right)$&3&2&$\frac{1}{6}$\\
\\
$(u^c)_L^i=$&$(u^c)_L$&$(c^c)_L$&$(t^c)_L$&$\bar 3$&1&$-\frac{2}{3}$\\
\\
$(d^c)_L^i=$&$(d^c)_L$&$(s^c)_L$&$(b^c)_L$&$\bar 3$&1&$\frac{1}{3}$\\
\\
$L_L^i=$&$\left(\begin{array}{l}\nu_{eL}\\e_L\end{array}\right)$
&$\left(\begin{array}{l}\nu_{\mu L}\\\mu_L\end{array}\right)$
&$\left(\begin{array}{l}\nu_{\tau
L}\\\tau_L\end{array}\right)$&1&2&$-\frac{1}{2}$\\
\\
$(e^c)_L^i=$&$(e^c)_L$&$(\mu^c)_L$&$(\tau^c)_L$&1&1&$1$
\end{tabular}
\caption{The fermion fields of the standard model and their gauge
quantum numbers.}
\end{center} \label{tab:sm}
\end{table}

The Lagrangian of the standard model is the sum of the gauge,
matter, Yukawa, and Higgs interactions,
\begin{equation}
{\cal L}_{SM} = {\cal L}_{Gauge} + {\cal L}_{Matter} + {\cal
L}_{Yukawa} + {\cal L}_{Higgs}\;.
\end{equation}
The pure gauge interactions contain the kinetic energies of the
gauge bosons and their self-interactions.  The ``matter''
Lagrangian contains the kinetic energy and gauge interactions of
the fermion fields,
\begin{equation}
{\cal L}_{Matter}=i\bar
Q_L^i\not\!\!DQ_L^i+i\overline{(u^c)}_L^i\not\!\!D(u^c)_L^i
+i\overline{(d^c)}_L^i\not\!\!D(d^c)_L^i+i\bar
L_L^i\not\!\!DL_L^i+i\overline{(e^c)}_L^i\not\!\!D(e^c)_L^i\;.
\end{equation}
A sum on the index $i$, which represents the generation, is
implied in the Lagrangian. To put this into the canonical form in
terms of left- and right-chiral fields, we use
\begin{equation}
(\psi^c)_L=C\gamma^0\psi^*_R
\end{equation}
which yields
\begin{equation}
{\cal L}_{Matter}=i\bar Q_L^i\not\!\!DQ_L^i+i\bar
u_R^i\not\!\!Du_R^i+i\bar d_R^i\not\!\!Dd_R^i+i\bar
L_L^i\not\!\!DL_L^i+i\bar e_R^i\not\!\!De_R^i\;.
\label{eq:Lmatter}
\end{equation}

At this stage, all the fermions are massless.  Majorana masses are
forbidden by the fact that all fermions carry hypercharge; in
addition, some transform under a complex representation of
$SU(3)$, and some transform under a pseudoreal representation of
$SU(2)_L$. Dirac masses are forbidden by the fact that no fermion
transforms under the complex-conjugate representation of another
fermion.

The absence of fermion masses implies that ${\cal L}_{Matter}$ has
a good deal of (accidental) global symmetry,
\begin{eqnarray*}
Q_L^i &\to &U_{Q_L}^{ij}Q_L^j\\
u_R^i &\to &U_{u_R}^{ij}u_R^j\\
d_R^i &\to &U_{d_R}^{ij}d_R^j\\
L_L^i &\to &U_{L_L}^{ij}L_L^j\\
e_R^i &\to &U_{e_R}^{ij}e_R^j\;.
\end{eqnarray*}
This symmetry is accidental in the sense that it is not imposed,
but rather follows from the fermion content and gauge symmetries
of the standard model.  Since there are five independent $U(3)$
symmetries, the global flavor symmetry of the matter Lagrangian is
$[U(3)]^5$.

These global flavor symmetries are violated by the Yukawa
couplings of the fermions to the Higgs field (see
Table~\ref{tab:higgs}),
\begin{equation}
{\cal L}_{Yukawa} = -\Gamma_u^{ij}\bar Q_L^i\epsilon
\phi^*u_R^j-\Gamma_d^{ij}\bar Q_L^i\phi d_R^j-\Gamma_e^{ij}\bar
L_L^i\phi e_R^j + h.c. \label{eq:LYukawa}
\end{equation}
where $\Gamma_u,\Gamma_d,\Gamma_e$ are $3\times 3$ complex
matrices in generation space.
\\[7pt]
{\em Exercise 2.1 - Show that if $\phi$ is an $SU(2)_L$ doublet,
then so is $\epsilon\phi^*$ (see Exercise 1.5).}
\\[7pt]
Only a very small subgroup of $[U(3)]^5$ is not violated,
corresponding to baryon number
\begin{eqnarray*}
Q_L^i &\to &e^{i\theta/3}Q_L^i \\
u_R^i &\to &e^{i\theta/3}u_R^i \\
d_R^i &\to &e^{i\theta/3}d_R^i
\end{eqnarray*}
and lepton number
\begin{eqnarray*}
L_L^i &\to &e^{i\phi}L_L^i \\
e_R^i &\to &e^{i\phi}e_R^i \;.
\end{eqnarray*}
Thus baryon number and lepton number are accidental global
symmetries of the standard model (see also Exercise 2.2).

\begin{table}[tb]
\begin{center}\begin{tabular}[7]{cccc}
&$\underline{SU(3)}$&$\underline{SU(2)_L}$&$\underline{U(1)_Y}$\\
\\
$\phi=\left(\begin{array}{l}\phi^+\\\phi^0\end{array}\right)$
&1&2&$\frac{1}{2}$
\end{tabular}
\caption{The Higgs field and its gauge quantum numbers.}
\end{center} \label{tab:higgs}
\end{table}

When the Higgs field acquires a vacuum expectation value,
\begin{equation}
\langle\phi\rangle=\left(\begin{array}{c}0\\
v/\sqrt 2\end{array}\right)\label{eq:vev}
\end{equation}
the fermion fields (except neutrinos) become massive via their
Yukawa couplings to the Higgs field, Eq.~(\ref{eq:LYukawa}),
\begin{equation}
{\cal L}_{M} = -M_u^{ij}\bar u_L^iu_R^j-M_d^{ij}\bar
d_L^id_R^j-M_e^{ij}\bar e_L^ie_R^j + h.c.\;, \label{eq:Lmass}
\end{equation}
where
\begin{equation}
M^{ij}=\Gamma^{ij}\frac{v}{\sqrt 2} \label{eq:massmatrix}
\end{equation}
are fermion mass matrices.  Thus $\psi_L$ and $\psi_R =
C\gamma^0(\psi^c)_L^*$ have paired up to make Dirac masses for
$u^i,d^i,e^i$. The neutrino field $\nu_L$ carries no unbroken
gauge symmetry, so it could potentially acquire a Majorana mass
\begin{equation}
{\cal L}=-{1\over 2}M_\nu^{ij}(\nu_L^{iT}C\nu_L^j + h.c.)\;.
\end{equation}
However, this term is forbidden by the accidental lepton number
symmetry.  Actually, baryon number and lepton number are
anomalous, but $B-L$ is not \cite{Peskin:1995ev}.  So it is more
precise to say that a Majorana neutrino mass is forbidden by
$B-L$.

The Yukawa matrices $\Gamma$ in Eq.~(\ref{eq:LYukawa}) are
$3\times 3$ complex matrices, and since there are three of them we
have apparently introduced $3\times 3\times 3\times 2=54$ new
parameters into the theory.  However, we will now show that only a
subset of these parameters are physically relevant.

Given this Lagrangian, one can proceed to calculate any physical
process of interest.  However, it is convenient to first perform
field redefinitions to make the physical content of the theory
manifest.  These field redefinitions do not change the predictions
of the theory; they are analogous to a change of variables when
performing an integration.  To make the masses of the fermions
manifest, we perform unitary field redefinitions on the fields in
order to diagonalize the mass matrices in Eq.~(\ref{eq:Lmass}):
\begin{eqnarray}
&u_L^i = A_{u_L}^{ij}u_L'^j & u_R^i = A_{u_R}^{ij}u_R'^j\nonumber\\
&d_L^i = A_{d_L}^{ij}d_L'^j & d_R^i = A_{d_R}^{ij}d_R'^j\nonumber\\
&e_L^i = A_{e_L}^{ij}e_L'^j & e_R^i = A_{e_R}^{ij}e_R'^j\nonumber\\
&\nu_L^i = A_{\nu_L}^{ij}\nu_L'^j\;. &
\end{eqnarray}
Each matrix $A$ must be unitary in order to preserve the form of
the kinetic-energy terms in the matter Lagrangian,
Eq.~(\ref{eq:Lmatter}), {\it e.g.}
\begin{equation}
{\cal L}_{KE} = i\bar u_L\not\!\partial u_L = i\bar
u'_LA_{u_L}^\dagger\not\!\partial A_{u_L}u'_L=i\bar
u'_L\not\!\partial u'_L
\end{equation}
where I have switched to index-free notation. Once the mass
matrices are diagonalized, the masses of the fermions are
manifest.  These transformations also diagonalize the Yukawa
matrices $\Gamma$, since they are proportional to the mass
matrices [see Eq.~(\ref{eq:massmatrix})]. However, we must
consider what effect these field redefinitions have on the rest of
the Lagrangian. They have no effect on the pure gauge or Higgs
parts of the Lagrangian, which are independent of the fermion
fields. They do affect the matter part of the Lagrangian,
Eq.~(\ref{eq:Lmatter}). However, a subset of these field
redefinitions is the global $[U(3)]^5$ symmetry of the matter
Lagrangian; this subset therefore has no effect.

One can count how many physically relevant parameters remain after
the field redefinitions are performed \cite{Falk:2000tx}.  Let's
concentrate on the quark sector.  The number of parameters
contained in the complex matrices $\Gamma_u,\Gamma_d$ is $2\times
3\times 3\times 2=36$. The unitary symmetries
$U_{Q_L},U_{u_R},U_{d_R}$ are a subset of the quark field
redefinitions; this subset will not affect the matter part of the
Lagrangian.  There are $3\times 3\times 3$ degrees of freedom in
these symmetries (a unitary $N\times N$ matrix has $N^2$ free
parameters), so the total number of parameters that remain in the
full Lagrangian after field redefinitions is
\begin{equation}
2\times 3\times 3\times 2 - (3\times 3\times 3-1) = 10
\end{equation}
where I have subtracted baryon number from the subset of field
redefinitions that are symmetries of the matter Lagrangian. Baryon
number is a symmetry of the Yukawa Lagrangian,
Eq.~(\ref{eq:LYukawa}), and hence cannot be used to diagonalize
the mass matrices.  The ten remaining parameters correspond to the
six quark masses and the four parameters of the
Cabibbo-Kobayashi-Maskawa (CKM) matrix (three mixing angles and
one $CP$-violating phase).
\\[7pt]
{\em Exercise 2.2 - Do the same counting for the lepton sector.
Show that it yields just one parameter.  Argue that this can be
interpreted as $m_e,m_\mu,m_\tau$ and two additional global $U(1)$
symmetries.  What are they?}
\\[7pt]
\indent Thus far the neutrinos are massless.  But we know that
neutrinos have mass.  Why not extend the standard model to include
the field $N_R$ (it will be clear shortly why I resist labeling
the field $\nu_R$), and add the Yukawa interaction
\begin{equation}
{\cal L}_{Yukawa}=-\Gamma_\nu^{ij}\bar L_L^i\epsilon \phi^*N_R^j +
h.c.\;.\label{eq:YukawaNR}
\end{equation}
{\em Exercise 2.3 - Show that $N_R$ is sterile, that is, carries
no gauge quantum numbers, and carries lepton number $+1$.}
\\[7pt]
Then when the Higgs field acquires a vacuum expectation value, the
neutrinos gain a Dirac mass just like all the other fermions.
There are two shortcomings to this proposal:
\begin{itemize}

\item There is no explanation of why neutrinos are so much lighter
than all the other fermions.

\item Since $N_R$ is sterile, the gauge symmetry allows a Majorana
mass term
\begin{equation}
{\cal L}=-{1\over 2}M_R^{ij}N_R^{iT}CN_R^j+h.c.\;.\label{eq:MajNR}
\end{equation}
With this term present, the fields $\nu_L$ and $N_R$ do not pair
up to make a Dirac fermion.  Rather, there are two Majorana
neutrinos per generation, which is more than we need to describe
nature.
\end{itemize}
We will later show how the latter shortcoming can be turned into a
virtue that addresses the first shortcoming.

The Majorana mass term of Eq.~(\ref{eq:MajNR}) is forbidden if
lepton number is imposed as an exact symmetry.  In that case, the
neutrino would acquire a Dirac mass from the Yukawa interaction of
Eq.~(\ref{eq:YukawaNR}).  However, recall that lepton number is an
accidental symmetry of the standard model; there is no reason to
expect it to be an exact symmetry when we extend the standard
model.  Even if this scenario were realized in nature, we should
still regard it as physics beyond the standard model, because it
requires the introduction of two new features: the field $N_R$,
and the elevation of lepton number from an accidental to an exact
symmetry.

Rather than add the field $N_R$, let's regard the standard model
as a low-energy effective field theory, with a Lagrangian that is
an expansion in inverse powers of some large mass $M$
\cite{Weinberg:1995mt,Georgi:1994qn},
\begin{equation}
{\cal L}={\cal L}_{SM}+\frac{1}{M}{\rm dim}\; 5+\frac{1}{M^2}{\rm
dim}\; 6+\cdots\;.\label{eq:Lexpansion}
\end{equation}
The terms beyond the standard model represent operators of higher
and higher dimension, starting with dimension five, which is the
least suppressed. Given the field content and gauge symmetries of
the standard model, there is only one dimension-five operator,
\begin{equation}
{\cal L}_5 = \frac{c^{ij}}{M}L_L^{iT}\epsilon\phi C\phi^T\epsilon
L_L^j + h.c.\;.\label{eq:L5}
\end{equation}
{\em Exercise 2.4 - Show that ${\cal L}_5$ is $SU(2)_L\times
U(1)_Y$ invariant, and that $c^{ij}$ is a symmetric matrix.  Show
that ${\cal L}_5$ violates lepton number.}
\\[7pt]
When the Higgs field acquires a vacuum expectation value,
Eq.~(\ref{eq:vev}), this yields a Majorana mass for the neutrino,
\begin{equation}
{\cal L}_{Maj} = - \frac{c^{ij}}{2}\frac{v^2}{M}\nu_L^{iT}C\nu_L^j
+ h.c.\;.\label{eq:LMaj}
\end{equation}
The neutrino is allowed to acquire a Majorana mass because lepton
number is violated by ${\cal L}_5$.  Thus lepton number is only a
low-energy accidental symmetry, and is in general violated by
higher-dimension operators.

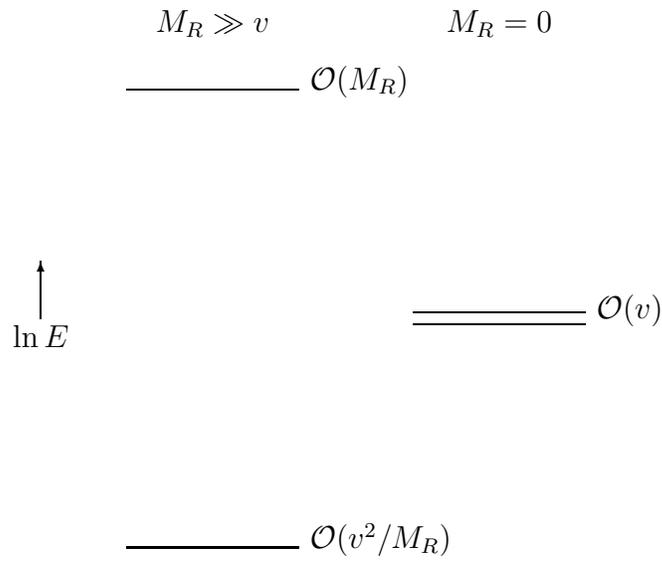
\begin{figure}

\setlength{\unitlength}{3in}
\begin{center}
\begin{picture}(1.2,1.2)
\put(0.05,0.5){\vector(0,1){0.1}}
\put(0.05,0.45){\makebox[0in][c]{$\ln E$}}
\put(0.2,0.1){\line(1,0){0.3}}
\put(0.52,0.1){\parbox[c]{1in}{$\mathcal O(v^2/M_R)$}}
\put(0.2,0.9){\line(1,0){0.3}}
\put(0.52,0.9){\parbox[c]{1in}{$\mathcal O(M_R)$}}
\put(0.7,0.49){\line(1,0){0.3}} \put(0.7,0.51){\line(1,0){0.3}}
\put(1.02,0.5){\parbox[c]{1in}{$\mathcal O(v)$}}
\put(0.35,1){\makebox[0in][c]{$M_R\gg v$}}
\put(0.85,1){\makebox[0in][c]{$M_R=0$}}
\end{picture}
\end{center}

\caption{If $M_R\gg v$, there is a heavy Majorana neutrino of mass
${\cal O}(M_R)$ and a light Majorana neutrino of mass ${\cal
O}(v^2/M_R)$.  If $M_R=0$, these neutrinos pair up to make a Dirac
neutrino of mass ${\cal O}(v)$.} \label{fig:seesaw}
\end{figure}

We see from Eq.~(\ref{eq:LMaj}) that neutrino masses are of order
$v^2/M$. Thus, if $M\gg v$, neutrino masses are naturally much
less than $v$. This is an attractive explanation of why neutrinos
are so much lighter than all the other fermions.
\\[7pt]
{\em Exercise 2.5 - Show that the MNS matrix (the analogue of the
CKM matrix in the lepton sector) has six physically relevant
parameters.  [Note: $c^{ij}$ is a symmetric, complex matrix.]}
\\[7pt]
\indent Let's now return to our discussion of the sterile
neutrino, $N_R^i$, and show how it yields Eq.~(\ref{eq:LMaj}) if
the sterile neutrino is made very heavy.  Begin with the
Lagrangians of Eqs.~(\ref{eq:YukawaNR}) and (\ref{eq:MajNR}),
\begin{equation}
{\cal L}=-\bar L_L\Gamma_\nu\epsilon \phi^*N_R -{1\over
2}N_R^{T}M_RCN_R+h.c.\;,\label{eq:LNR}
\end{equation}
where I have switched to an index-free notation. If $M_R$ is very
large, then the field $N_R$ is not present at low energy, and we
can integrate it out of the theory. At low energy, $N_R$ acts like
a non-dynamical, classical field, and we can remove it by solving
its equation of motion,
\begin{equation}
\frac{\partial{\cal L}}{\partial N_R} = 0\;,
\end{equation}
and plugging it back into Eq.~(\ref{eq:LNR}).  This yields
Eq.~(\ref{eq:LMaj}), where
\begin{equation}
\frac{c}{M} = -{1\over 2}(\Gamma_\nu
M_R^{-1}\Gamma_\nu^T)^\dagger\;.\label{eq:cM}
\end{equation}
Thus we see that the mass $M$ in our effective-field-theory
description, Eq.~(\ref{eq:Lexpansion}), is proportional to the
mass $M_R$. The effective field theory breaks down when one
approaches energy of order the mass of the heavy Majorana neutrino
$N_R$.
\\[7pt]
{\em Exercise 2.6 - Derive Eq.~(\ref{eq:cM}).  [Hint: If you
follow the above steps, you'll arrive at the $h.c.$ term of ${\cal
L}_5$.]}
\\[7pt]
\indent I show in Fig.~\ref{fig:seesaw} the spectrum of neutrino
masses of a single generation in the two limits we have just
discussed. If $M_R\gg v$, then $N_R$ is a heavy Majorana neutrino
of mass $M_R$, and $\nu_L$ is a very light Majorana neutrino of
mass ${\cal O}(v^2/M_R)$.  In the other extreme, if $M_R=0$ (which
would be the case if lepton number were an exact symmetry),
$\nu_L$ and $N_R$ pair up to make a Dirac neutrino of mass ${\cal
O}(v)$. This exemplifies the fact that a Dirac neutrino can be
thought of as two degenerate Majorana neutrinos (of opposite
charge conjugation), which we showed at the end of the previous
section.

A heavy neutrino of mass $M_R$ yielding a light neutrino of mass
${\cal O}(v^2/M_R)$ is called the ``seesaw'' mechanism: compared
with the mass scale $v$, one has gone up and the other has gone
down \cite{Kayser:1989iu}.  The effective-field-theory
description, Eq.~(\ref{eq:Lexpansion}), can be regarded as a
generalization of the see-saw mechanism, one that does not depend
on the details of the new physics residing at the scale $M$.

\newpage

\section{Custodial symmetry}
\label{sec:custodial}

We now leave the flavor sector of the standard model to begin a
discussion of the Higgs sector, or more generally, the
electroweak-symmetry-breaking sector.  In this section we consider
the global symmetries of this sector.

The Higgs sector of the standard model is described by the
Lagrangian
\begin{equation}
{\cal L}_{Higgs}=(D_\mu\phi)^\dagger
D^\mu\phi-V(\phi)\label{eq:Higgs}
\end{equation}
where the Higgs potential is
\begin{equation}
V(\phi)=-\mu^2\phi^\dagger\phi+\lambda
(\phi^\dagger\phi)^2\label{eq:V}
\end{equation}
and the gauge-covariant derivative is
\begin{equation}
D_\mu\phi=\left(\partial_\mu+i\frac{g}{2}\sigma\cdot
W_\mu+i\frac{g'}{2}B_\mu\right)\phi\;.
\end{equation}
The Higgs Lagrangian is $SU(2)_L\times U(1)_Y$ symmetric by
construction, but it also has an approximate (accidental) global
symmetry.  To see this, it is useful to rewrite the Lagrangian as
follows. Label the components of the Higgs-doublet field as
\begin{equation}
\phi=\left(\begin{array}{c}\phi^+\\
\phi^0\end{array}\right)\;.
\end{equation}
Then $\epsilon\phi^*$, which is also an $SU(2)_L$ doublet (see
Exercise 2.1), has components
\begin{equation}
\epsilon\phi^*=\left(\begin{array}{c}\phi^{0*}\\
-\phi^-\end{array}\right)
\end{equation}
where I've used $\phi^-=\phi^{+*}$.  Now define a Higgs matrix (or
bi-doublet) field
\begin{equation}
\Phi=\frac{1}{\sqrt 2}(\epsilon\phi^*,\phi)
=\frac{1}{\sqrt 2}\left(\begin{array}{cc}\phi^{0*}&\phi^+\\
-\phi^-&\phi^0\end{array}\right)\;.
\end{equation}
We can rewrite the Higgs Lagrangian in terms of this matrix field
as
\begin{equation}
{\cal L}_{Higgs}={\rm Tr}\; (D_\mu\Phi)^\dagger
D^\mu\Phi-V(\Phi)\label{eq:Higgs2}
\end{equation}
where the potential is given by
\begin{equation}
V(\Phi)=-\mu^2{\rm Tr}\; \Phi^\dagger \Phi + \lambda ({\rm Tr}\;
\Phi^\dagger \Phi)^2
\end{equation}
and the gauge-covariant derivative by
\begin{equation}
D_\mu \Phi = \left(\partial_\mu\Phi+i\frac{g}{2}\sigma\cdot
W_\mu\Phi-i\frac{g'}{2}B_\mu\Phi\sigma_3\right)\;.
\end{equation}
Note the Pauli matrix $\sigma_3$ at the end of the last term in
the above equation.  This is necessary because $\phi$ and
$\epsilon\phi^*$ have opposite hypercharge.
\\[7pt]
{\em Exercise 3.1 - Verify that Eqs.~(\ref{eq:Higgs}) and
(\ref{eq:Higgs2}) are identical.}
\\[7pt]
\indent The $SU(2)_L\times U(1)_Y$ gauge symmetry acts on the
Higgs matrix field as
\begin{eqnarray}
SU(2)_L: && \Phi \to L\Phi \nonumber\\
U(1)_Y: && \Phi \to \Phi e^{-{i\over
2}\sigma_3\theta}\label{eq:gauge}
\end{eqnarray}
where the $\sigma_3$ in the hypercharge transformation is again
due to the opposite hypercharges of $\phi$ and $\epsilon\phi^*$.
Under $SU(2)_L$,
\begin{equation}
{\rm Tr}\; (D_\mu\Phi)^\dagger D^\mu\Phi\to {\rm Tr}\;
(D_\mu\Phi)^\dagger L^\dagger LD^\mu\Phi={\rm Tr}\;
(D_\mu\Phi)^\dagger D^\mu\Phi
\end{equation}
which shows it is invariant.

To make the approximate global symmetry manifest, take the limit
that the hypercharge coupling vanishes, $g'\to 0$.  The Higgs
Lagrangian is still given by Eq.~(\ref{eq:Higgs2}), but now the
gauge-covariant derivative is given simply by
\begin{equation}
D_\mu \Phi = \left(\partial_\mu+i\frac{g}{2}\sigma\cdot
W_\mu\right)\Phi\;.
\end{equation}
We see that in this limit the Lagrangian has a global symmetry
$SU(2)_R$, given by
\begin{eqnarray}
SU(2)_R: && \Phi \to \Phi R^\dagger\;.\label{eq:SU2R}
\end{eqnarray}
Under an $SU(2)_R$ transformation,
\begin{equation}
{\rm Tr}\; (D_\mu\Phi)^\dagger D^\mu\Phi\to {\rm Tr}\;
R(D_\mu\Phi)^\dagger D^\mu\Phi R^\dagger={\rm Tr}\;
(D_\mu\Phi)^\dagger D^\mu\Phi
\end{equation}
which shows it is invariant.  Thus in the limit $g'\to 0$, the
Higgs sector of the standard model has the global symmetry
$SU(2)_L\times SU(2)_R$, where $SU(2)_L$ is just the global
version of the gauge symmetry, and $SU(2)_R$ is an approximate,
accidental global symmetry:
\begin{eqnarray}
SU(2)_L\times SU(2)_R: && \Phi \to L\Phi R^\dagger\;.
\end{eqnarray}
{\em Exercise 3.2 - Show that $U(1)_Y$ is a subgroup of
$SU(2)_R$.}
\\[7pt]
\indent When the Higgs field acquires a vacuum expectation value,
Eq.~(\ref{eq:vev}), the matrix field is
\begin{equation}
\langle\Phi\rangle
=\frac{1}{2}\left(\begin{array}{cc}v&0\\
0&v\end{array}\right)\;.
\end{equation}
This breaks both $SU(2)_L$ and $SU(2)_R$,
\begin{equation}
L\langle\Phi\rangle\neq\langle\Phi\rangle\;\;\;\;\langle\Phi\rangle
R^\dagger\neq\langle\Phi\rangle
\end{equation}
but leaves unbroken the subgroup $SU(2)_{L+R}$, corresponding to
simultaneous $SU(2)_L$ and $SU(2)_R$ transformations with $L=R$:
\begin{equation}
L\langle\Phi\rangle L^\dagger=\langle\Phi\rangle\;.
\end{equation}
Thus the Higgs vacuum expectation value breaks the global symmetry
in the pattern
\begin{equation}
SU(2)_L\times SU(2)_R \to SU(2)_{L+R}\;.
\end{equation}
This is called ``custodial symmetry;'' actually, some authors
refer to $SU(2)_R$ by this name \cite{Georgi:1994qn}, while others
reserve it for $SU(2)_{L+R}$ \cite{Sikivie:1980hm}.

Since $SU(2)$ is a three-dimensional group, the number of broken
generators is $3+3-3=3$. These give rise to three massless
Goldstone bosons, which are then eaten by the Higgs mechanism to
provide the mass of the $W^+$, $W^-$, and $Z$ bosons,
\begin{eqnarray}
M_W^2 &=& {1\over 4}g^2v^2\nonumber \\
M_Z^2 &=& {1\over 4}(g^2+g'^2)v^2\;.\label{eq:WZmass}
\end{eqnarray}
Thus
\begin{equation}
\frac{M_W^2}{M_Z^2}=\frac{g^2}{g^2+g'^2}=\cos^2\theta_W
\end{equation}
or
\begin{equation}
\rho=\frac{M_W^2}{M_Z^2\cos^2\theta_W}=1 \label{eq:rho}
\end{equation}
at tree level.
\\[7pt]
{\em Exercise 3.3 - Show that $W_\mu^A$ transforms as a triplet
under global $SU(2)_L$ and a singlet under $SU(2)_R$, and hence as
a triplet under the unbroken $SU(2)_{L+R}$.}
\\[7pt]
Thus, in the limit $g'\to 0$, $W^+,W^-,Z$ form a triplet of an
unbroken global symmetry.  This explains why $M_W=M_Z$ in the
$g'\to 0$ limit.

\begin{figure}[b]
\begin{center}
\epsfxsize=4.5in \epsfbox{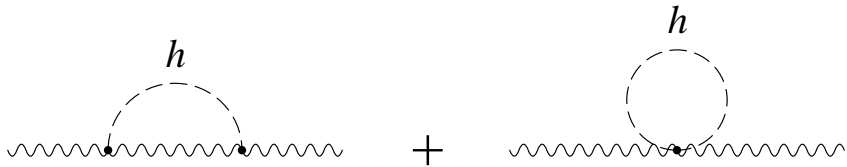}
\end{center}
\caption{Virtual Higgs-boson loops contribute to the $W$ and $Z$
masses.} \label{fig:higgsloops}
\end{figure}

Custodial symmetry also helps us understand properties of the
theory beyond tree level.  Due to the unbroken $SU(2)_{L+R}$ in
the $g'\to 0$ limit, radiative corrections to the $\rho$ parameter
in Eq.~(\ref{eq:rho}) due to gauge and Higgs bosons must be
proportional to $g'^2$. For example, the leading correction to the
$\rho$ parameter from loops of Higgs bosons
(Fig.~\ref{fig:higgsloops}) in the $\overline{MS}$ scheme is
\begin{equation}
\hat\rho\approx 1-\frac{11G_FM_Z^2\sin^2\theta_W}{24\sqrt
2\pi^2}\ln\frac{m_h^2}{M_Z^2}\;.
\end{equation}
This correction vanishes in the limit $g'\to 0$
($\sin^2\theta_W\to 0$).  The custodial symmetry protects the
tree-level relation $\rho=1$ from radiative corrections, and hence
it's name.  This leading correction, proportional to $\ln m_h$,
allows us to bound the Higgs-boson mass from precision electroweak
measurements.

\begin{figure}[t]
\begin{center}
\epsfxsize=4.5in \epsfbox{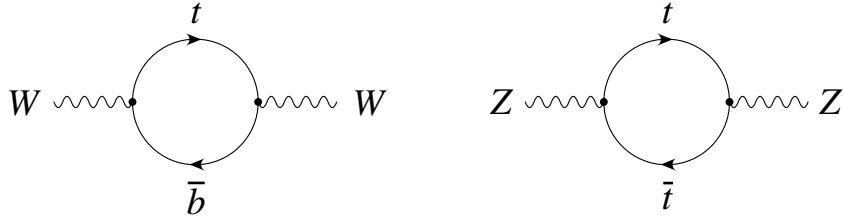}
\end{center}
\caption{Virtual top-quark loops contribute to the $W$ and $Z$
masses.} \label{fig:toploops}
\end{figure}

Custodial symmetry also helps us understand radiative corrections
due to massive fermions, as shown in Fig.~\ref{fig:toploops}.  The
leading correction due to loops of top and bottom quarks is
\cite{Veltman:1994wz}
\begin{equation}
\hat\rho\approx 1+\frac{3G_F}{8\pi^2\sqrt
2}\left(m_t^2+m_b^2-2\frac{m_t^2m_b^2}{m_t^2-m_b^2}\ln\frac{m_t^2}{m_b^2}\right)\;.
\end{equation}
\\[7pt]
{\em Exercise 3.4 - Show that this correction vanishes in the
limit $m_t=m_b$.}
\\[7pt]
{\em Exercise 3.5 - Show that the $t,b$ Yukawa couplings have a
custodial symmetry in the limit $m_t=m_b$.}
\\[7pt]
This leading correction, proportional to the square of the fermion
mass, allowed us to predict the top-quark mass from precision
electroweak measurements before it was discovered.

Thus we see that custodial symmetry is vital to our understanding
of the electroweak sector.  However, the physical Higgs boson
itself is not really necessary.  As long as the
electroweak-symmetry-breaking mechanism possesses custodial
symmetry, the $\rho$ parameter equals unity at tree level and is
protected from large radiative corrections.

Let's develop an effective field theory of electroweak symmetry
breaking that makes custodial symmetry manifest
\cite{Chivukula:1995hr}.  A simple way to do this is to replace
the matrix field $\Phi$ with another matrix field, $\Sigma$, which
contains the Goldstone bosons, $\pi_i$ (which are eaten by the
weak vector bosons), but does not contain a physical Higgs boson:
\begin{equation}
\Phi\to
\frac{v}{2}\Sigma\;\;\;\;\;\;\Sigma=e^{i\frac{\sigma\cdot\pi}{v}}\;.
\end{equation}
The Lagrangian for electroweak symmetry breaking (EWSB) is the
analogue of Eq.~(\ref{eq:Higgs2}),
\begin{equation}
{\cal L}_{EWSB}=\frac{v^2}{4}{\rm Tr}\;(D_\mu\Sigma)^\dagger
D^\mu\Sigma\;.\label{eq:EWSB}
\end{equation}
The Goldstone-boson matrix field $\Sigma$ transforms under
custodial symmetry as
\begin{eqnarray*}
SU(2)_L: && \Sigma\to L\Sigma\\
SU(2)_R: && \Sigma \to \Sigma R^\dagger\\
SU(2)_{L+R}: && \Sigma \to L\Sigma L^\dagger\;.
\end{eqnarray*}
\\[7pt]
{\em Exercise 3.6 - Let $L=e^{\frac{i}{2}\sigma\cdot\theta}$ be an
$SU(2)_L$ transformation.  Show that under an infinitesimal
$SU(2)_L$ transformation, $\pi_i$ transforms non-linearly,
\begin{equation}
\pi_i\to \frac{v}{2}\theta_i+\pi_i
\end{equation}
while under an infinitesimal $SU(2)_{L+R}$ transformation it
transforms linearly,
\begin{equation}
\pi_i\to \pi_i-\epsilon_{ijk}\theta_j\pi_k\;.
\end{equation}}
\\[7pt]
Since $SU(2)_L\times SU(2)_R$ is realized non-linearly, this model
is called the non-linear sigma model.  The fact that the symmetry
is realized non-linearly means that it is broken.

Unitary gauge corresponds to $\Sigma=1$.  The
electroweak-symmetry-breaking Lagrangian, Eq.~(\ref{eq:EWSB}),
yields the correct vector boson masses of Eq.~(\ref{eq:WZmass}).
The only other dimension-two operator allowed by the
$SU(2)_L\times U(1)_Y$ gauge symmetry is
\begin{equation}
{\cal L}=c\frac{v^2}{4}[{\rm Tr}\;\sigma_3\Sigma^\dagger
D_\mu\Sigma]^2\;.
\end{equation}
This term contributes to the $W$ and $Z$ masses such that
\begin{equation}
\rho = 1 + 2c\;.
\end{equation}
{\em Exercise 3.7 - Show that this operator violates the custodial
symmetry.}
\\[7pt]
This proves that custodial symmetry is sufficient to produce
$\rho=1$ at leading order in the effective theory of electroweak
symmetry breaking.

Beyond leading order, the effective theory has a tower of
higher-dimension operators,
\begin{equation}
{\cal L}=\frac{v^2}{4}{\rm Tr}\;(D_\mu\Sigma)^\dagger
D^\mu\Sigma+{\rm dim}\; 4+\cdots\;.
\end{equation}
The power counting is different from that of the effective field
theory we previously encountered in Eq.~(\ref{eq:Lexpansion}),
where the leading term, ${\cal L}_{SM}$, is renormalizable.  Here
the leading term, of dimension two, is non-renormalizable.  The
effects of the higher-dimension operators are suppressed by
$E^2/\Lambda^2$, where $\Lambda \le 4\pi v$.

All we are really sure of is that the electroweak symmetry is
broken and that the electroweak symmetry breaking sector has an
approximate custodial symmetry.  We don't really know if a Higgs
boson exists.  The Fermilab Tevatron and the CERN Large Hadron
Collider will settle that question.

\newpage

\section{Scale symmetry}
\label{sec:scale}

Although we do not really know the mechanism of electroweak
symmetry breaking, let alone whether a Higgs boson exists, it is
appropriate to consider the minimal model with a single Higgs
doublet as the ``standard model.''  This is due to the remarkable
fact that precision electroweak measurements are consistent with a
relatively light Higgs boson, $m_h=114^{+69}_{-45}$ GeV
\cite{Eidelman:2004wy}.  It could have turned out that these
measurements were not consistent with a Higgs boson of any mass,
but that is not the way things have unfolded.

As discussed in Section~\ref{sec:flavor}, it is likely that the
standard model is the leading term in an effective Lagrangian,
Eq.~(\ref{eq:Lexpansion}), corresponding to an expansion in
inverse powers of some large scale $M$.  There are at least three
hints that the scale $M$ is very large, much greater than the weak
scale, $v$:
\begin{itemize}
\item As discussed in Section~\ref{sec:flavor}, Majorana neutrino
masses are of order $v^2/M$, which implies $M\sim 10^{14}-10^{16}$
GeV.
\item Attempts at grand unification indicate that unification of
the gauge couplings of the standard model occurs at around
$10^{16}$ GeV.
\item Quantum gravity becomes important at or before the Planck
scale, $M_{Pl}=(\hbar c/G_N)^{1/2}\sim 10^{19}$ GeV.
\end{itemize}
This raises a puzzle: if the fundamental scale of physics, $M$, is
so high, why does the standard model reside at the ordinary
energies that we observe, rather than at $M$?  The standard model
explains this in part, but not entirely.

Let's begin with fermions.  As we discussed in
Section~\ref{sec:flavor}, the fermions transform under a complex
representation of the gauge symmetry.  There are no fermions that
are gauge singlets or transform under a real representation, so
Majorana masses are forbidden.  There are no pairs of fermions
which transform under complex-conjugate representations of the
gauge symmetry, so Dirac masses are forbidden.  Thus fermions are
massless until the $SU(2)_L\times U(1)_Y$ symmetry is broken.
Hence fermion masses are naturally of ${\cal O}(v)$.

Although this is a successful explanation of why fermions are
light compared with $M$, it is not entirely satisfactory.  Only
the top quark has a mass of ${\cal O}(v)$; all the other charged
fermions are considerably lighter.  In the standard model, this is
due to the very small Yukawa couplings of the fermions.  It is
puzzling that these couplings are so small; for the electron, the
Yukawa coupling is about $10^{-5}$.  These small couplings suggest
that there is an approximate flavor symmetry at work (see
Section~\ref{sec:flavor}), only weakly violated by the Yukawa
couplings.

As discussed in Section~\ref{sec:flavor}, neutrino masses are zero
at leading order in the effective theory due to an accidental
lepton number symmetry. This symmetry is violated by the dimension
5 operator in Eq.~(\ref{eq:L5}), giving rise to a small neutrino
mass of ${\cal O}(v^2/M)$.  As mentioned above, this is one of the
reasons we believe that $M\gg v$.

In addition to the fermion masses, the flavor sector also contains
the CKM and MNS mixing matrices, whose mixing angles are listed in
Table~\ref{tab:mixing}. The standard model accommodates, but does
not explain, the pattern of masses and mixings.  Hidden in this
pattern are clues to physics beyond the standard model.

\begin{table}[tb]
\begin{center}
\begin{tabular}{|c||c|c|}
\hline
angle & quark & lepton \\
\hline $\theta_{12}$ & $13^\circ$ & $34^\circ$ \\
$\theta_{23}$ & $2.3^\circ$ & $45^\circ$ \\
$\theta_{13}$ & $0.23^\circ$ & $\le 12^\circ$ \\
$\delta$ & $60^\circ$ & unknown \\
\hline \hline
\end{tabular}
\caption{The observed mixing angles in the quark and lepton
sectors.  If neutrinos are Majorana, there are three unknown
$CP$-violating phases; if Dirac, only one.}
\end{center}
\label{tab:mixing}
\end{table}

Let's now turn to the gauge bosons.  Gauge bosons are associated
with a local (gauge) symmetry that protects their masslessness.
Under a gauge transformation, the gauge bosons transform as
\begin{equation}
T\cdot A_\mu \to UT\cdot A_\mu U^{-1}+\frac{i}{g}(\partial_\mu
U)U^\dagger
\end{equation}
where $T^A$ are the generators of the group.  It is the second
term in this transformation that forbids a mass term,
\begin{equation}
{\cal L}={1\over 2}M^2A_\mu\cdot A^\mu = M^2 {\rm Tr}\;T\cdot
A_\mu T\cdot A^\mu
\end{equation}
where ${\rm Tr}\;T^AT^B={1\over 2}\delta^{AB}$.  Thus gauge bosons
are massless unless the gauge symmetry is broken. Hence the photon
and gluons are massless, while the weak vector bosons naturally
have a mass of ${\cal O}(v)$.

Gauge bosons are associated with a local symmetry, but it is not
clear which way the logic runs: are they massless because of the
local symmetry, or is the local symmetry present because they are
massless?  It's not even clear that it makes sense to think about
it either of these ways.  For many years physicists considered the
local symmetry to be fundamental, but that position is no longer
held sacred. For example, in string theory there are massless,
spin-one modes of the string, and at low energy these are
described by an effective gauge theory, even though the underlying
string theory is not based on gauge symmetry
\cite{Weinberg:1995mt}. Perhaps even more remarkable are
supersymmetric field theories with electric-magnetic duality.
There are examples in which the long-distance (magnetic) degrees
of freedom include massless gauge bosons with a local symmetry,
yet the short-distance (electric) degrees of freedom do not
possess this local symmetry \cite{Schwarz:1998ny}.

Because of its importance, I want to dwell in more detail on the
connection between massless gauge bosons and gauge symmetry. For
simplicity, I'll specialize to QED, although the results for a
non-Abelian gauge theory are analogous.

\begin{figure}[t]
\begin{center}
\epsfxsize=2in \epsfbox{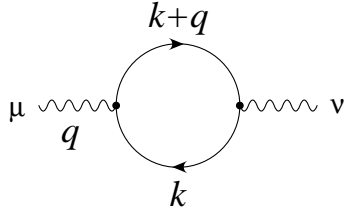}
\end{center}
\caption{Photon self-energy diagram.} \label{fig:Pi}
\end{figure}

\begin{figure}[b]
\begin{center}
\epsfxsize=6in \epsfbox{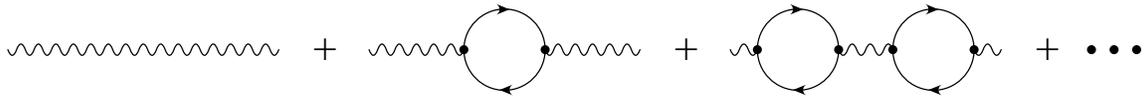}
\end{center}
\caption{The renormalized photon propagator from a sum of
self-energy diagrams.} \label{fig:Dyson}
\end{figure}

Gauge symmetry implies a Ward identity. In the case of the photon
self energy $\Pi_{\mu\nu}$, shown in Fig.~\ref{fig:Pi}, the Ward
identity is
\begin{equation}
q^\mu\Pi_{\mu\nu}=0\label{eq:Ward}
\end{equation}
where $q$ is the photon four momentum.  This implies that the self
energy must have the form
\begin{equation}
\Pi_{\mu\nu}=(q^2g_{\mu\nu}-q_\mu q_\nu)\Pi(q^2)\;.\label{eq:self}
\end{equation}
We can calculate the effect of the photon self energy on the
photon propagator by summing the geometric series shown in
Fig.~\ref{fig:Dyson}. It is particularly easy to do this in Landau
gauge, where the numerator of the propagator has the same form as
the self energy, Eq.~(\ref{eq:self}).  We find
\begin{eqnarray}
&&\frac{-i}{q^2}\left(g^{\mu\nu}-\frac{q^\mu q^\nu}{q^2}\right)
+\frac{-i}{q^2}\left(g^{\mu\rho}-\frac{q^\mu q^\rho}{q^2}\right)
i(q^2g_{\rho\sigma}-q_\rho q_\sigma)\Pi
\frac{-i}{q^2}\left(g^{\sigma\nu}-\frac{q^\sigma
q^\nu}{q^2}\right)+\cdots\nonumber\\
&&=\frac{-i}{q^2}\left(g^{\mu\nu}-\frac{q^\mu
q^\nu}{q^2}\right)[1+\Pi+\cdots]\nonumber\\
&&=\frac{-i}{q^2[1-\Pi]}\left(g^{\mu\nu}-\frac{q^\mu
q^\nu}{q^2}\right)\;.\label{eq:propagator}
\end{eqnarray}
Thus the propagator acquires a factor $[1-\Pi]^{-1}$.

Let's now proceed to calculate the contribution to $\Pi$ from a
fermion loop.  Using the Feynman diagram in Fig.~\ref{fig:Pi}, we
find \cite{Peskin:1995ev}
\begin{equation}
i\Pi_{\mu\nu}=i(q^2g_{\mu\nu}-q_\mu q_\nu )\Pi
=(-ieQ)^2i^2(-1)\int\frac{d^Nk}{(2\pi)^N} \frac{{\rm
Tr}\;\gamma_\mu(\slashed k+m)\gamma_\nu(\slashed k+\slashed
q+m)}{(k^2-m^2)((k+q)^2-m^2)}
\end{equation}
where $Q$ is the fermion electric charge and the factor $(-1)$ is
from the fermion loop.  Let's calculate in $N$ dimensions, in
order to regulate the ultraviolet divergence.  Contracting both
sides of this equation with $g^{\mu\nu}$ yields
\begin{equation}
i(q^2N-q^2)\Pi =-e^2Q^2\int\frac{d^Nk}{(2\pi)^N} \frac{{\rm
Tr}\;\gamma_\mu(\slashed k+m)\gamma^\mu(\slashed k+\slashed
q+m)}{(k^2-m^2)((k+q)^2-m^2)}\label{eq:Pi}
\end{equation}
where I've used $g^{\mu\nu}g_{\mu\nu}=N$. The trace gives
\begin{equation}
{\rm Tr}\;\gamma_\mu(\slashed k+m)\gamma^\mu(\slashed k+\slashed
q+m)=4[-2(1-\epsilon)(k^2+k\cdot q)+(4-2\epsilon)m^2]
\end{equation}
where I've used
\begin{eqnarray*}
\gamma_\mu \gamma^\mu &=& N \equiv 4-2\epsilon\\
\gamma_\mu \slashed a\gamma^\mu &=& -2(1-\epsilon)\slashed a\;.
\end{eqnarray*}
Now let's rewrite the numerator in terms of factors appearing in
the denominator,
\begin{equation}
k^2+k\cdot q={1\over 2}[(k+q)^2-m^2+(k^2-m^2)-q^2+2m^2]\;.
\end{equation}
This allows us to write Eq.~(\ref{eq:Pi}) in the simple form
\begin{eqnarray}
&&i(3-2\epsilon)q^2\Pi=-4e^2Q^2\nonumber\\
&&\int\frac{d^Nk}{(2\pi)^N}
\left[(\epsilon-1)\left(\frac{1}{(k^2-m^2)}+\frac{1}{((k+q)^2-m^2)}\right)
+\frac{(1-\epsilon)q^2+2m^2}{(k^2-m^2)((k+q)^2-m^2)}\right]\;.\label{eq:Pi2}
\end{eqnarray}

Let's first work in $N=4$ dimensions ($\epsilon=0$).  The first
two integrals above are quadratically divergent.  Let's simply cut
them off. The first integral gives
\begin{equation}
\int\frac{d^4k}{(2\pi)^4}
\frac{1}{(k^2-m^2)}=i\int\frac{d^4k_E}{(2\pi)^4}
\frac{1}{(-k_E^2-m^2)}\sim -\frac{i}{16\pi^2}M^2
\end{equation}
where I've Wick rotated to Euclidean space ($k^0\to ik^0_E$) and
used $d^4k_E=\pi^2k_E^2dk_E^2$ (after angular integration) to
evaluate the integral, cut off at $k_E^2=M^2$. The second integral
gives the same result after first performing the shift $k\to k-q$.
Thus we find
\begin{equation}
\Pi\sim -\frac{e^2}{6\pi^2}Q^2\frac{M^2}{q^2}\;.
\end{equation}
Inserting this into Eq.~(\ref{eq:propagator}), we see that the
photon has acquired a (tachyonic) mass of ${\cal O}(M)$:
\begin{equation}
\frac{-i}{q^2[1-\Pi]}\sim
\frac{-i}{q^2+\frac{e^2}{6\pi^2}Q^2M^2}\;.
\end{equation}

What went wrong?  Although it is not evident from our calculation,
simply cutting off the integral violates the Ward identity,
Eq.~(\ref{eq:Ward}) \cite{Peskin:1995ev}:
\begin{equation}
\Pi_{\mu\nu}\sim M^2g_{\mu\nu} \to q^\mu\Pi_{\mu\nu}\neq 0\;.
\end{equation}
Thus the Ward identity, which follows from gauge symmetry, is
essential to protect the masslessness of the photon.

Rather than using a cutoff, we can evaluate the photon self energy
in $N$ dimensions, which respects the Ward identity.  The
quadratically divergent integrals give
\begin{equation}
\int\frac{d^Nk}{(2\pi)^N}
\frac{1}{(k^2-m^2)}=\int\frac{d^Nk}{(2\pi)^N}
\frac{1}{((k+q)^2-m^2)}=-\frac{i}{(4\pi)^{N/2}}\Gamma(\epsilon-1)(m^2)^{1-\epsilon}\;.
\end{equation}
The pole at $\epsilon=1$ ($N=2$) signals the quadratic divergence.
However, these integrals are multiplied by a factor $(\epsilon-1)$
in Eq.~(\ref{eq:Pi2}).  Using
\begin{equation}
(\epsilon-1)\Gamma(\epsilon-1)=\Gamma(\epsilon)
\end{equation}
we see that the pole at $\epsilon=1$, and hence the quadratic
divergence, was illusory.  The quadratic dependence on the cutoff
$M$ that we discovered above was an artifact of using a regulator
that violates the Ward identity.

After some additional work, one finds \cite{Peskin:1995ev}
\begin{equation}
\Pi=-\frac{8e^2Q^2}{(4\pi)^{N/2}}\Gamma(\epsilon)\int_0^1 dx
\frac{x(1-x)}{[m^2-x(1-x)q^2]^\epsilon}\;.
\end{equation}
The pole at $\epsilon=0$ ($N=4$) signals a logarithmic divergence.
Inserting this into Eq.~(\ref{eq:propagator}), we find that the
photon propagator still has a pole at $q^2=0$, so the photon
remains massless, thanks to the Ward identity.

Thus the standard model successfully explains why fermions and
gauge bosons are so much lighter than the hypothesized fundamental
scale, $M$.  Now let's turn to scalars.  Here is where the
standard model is not so successful.  A scalar mass term is of the
form
\begin{equation}
{\cal L}=-m^2\phi^\dagger\phi
\end{equation}
and this is always allowed regardless of the gauge symmetry.  Thus
there is no reason for scalars to be light compared to $M$.  In
the standard model, this would mean that the Higgs field would
naturally have a mass of ${\cal O}(M)$, and thus not be available
at low energy to break the electroweak symmetry.

There are at least two ways to avoid this conclusion. One way is
to make the scalars Goldstone bosons of some broken global
symmetry.  As we saw in Exercise 3.6, Goldstone bosons transform
as $\pi_i\to {v\over 2}\theta_i+\pi_i$, which forbids a mass term,
${\cal L}=-m^2\pi_i\pi_i$.  Perhaps the Higgs field is a
(pseudo)-Goldstone boson of some (approximate) broken global
symmetry.  This is the idea behind the currently popular ``little
Higgs'' models \cite{Schmaltz:2002wx}.

The other generic way to make scalars massless is to use
supersymmetry \cite{Martin:1997ns}.  A scalar field is replaced by
a scalar superfield, $\phi\to \Phi$ (not to be confused with the
matrix field $\Phi$ of Section~\ref{sec:custodial}). The
superpotential $W(\Phi)$ depends only on $\Phi$, not $\Phi^*$,
which is referred to as the property of holomorphy. If the scalar
field transforms under a complex or pseudoreal representation of
some global or local group, then a mass term, $W(\Phi)={1\over
2}m\Phi^2$, is forbidden, exactly as we saw for fermions in
Section~\ref{sec:flavor}.  Since supersymmetry connects bosons and
fermions, it is not surprising that the same method we used to
make fermions light can also be used to make scalars light.

Unfortunately, the minimal supersymmetric standard model does not
make use of this method.  That model requires two light
Higgs-doublet superfields, $H_1$ and $H_2$, with opposite
hypercharge, in order to provide masses to both up-type and
down-type quarks. Thus a mass term in the superpotential
\begin{equation}
W(H_1,H_2)=\mu H_2^T\epsilon H_1
\end{equation}
is allowed by the gauge symmetry.  However, this does not mean
that supersymmetry is irrelevant to understanding why the Higgs
scalars are light compared to $M$, as we will discuss shortly.

\begin{figure}[t]
\begin{center}
\epsfxsize=2in \epsfbox{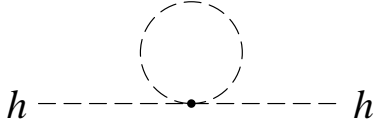}
\end{center}
\caption{Quadratically divergent contribution to the Higgs-boson
mass.} \label{fig:quad_div}
\end{figure}

We conclude that in the standard model, there is no principle to
tell us why the Higgs mass is light compared to the hypothesized
scale of fundamental physics, $M$.  It is not unreasonable to
regard this as a merely ``aesthetic'' problem.  However, there is
an even worse problem: it is not natural for a scalar mass to be
much lighter than $M$.  Consider the one-loop contribution to the
scalar mass shown in Fig.~\ref{fig:quad_div}.  This diagram has a
genuine quadratic divergence, so the relation between the Higgs
mass evaluated at low energy and at the scale $M$ is
\begin{equation}
m^2(m)=m^2(M)+\frac{3}{4\pi^2}\lambda M^2
\end{equation}
where $\lambda$ is the Higgs-field self coupling,
Eq.~(\ref{eq:V}). In order to obtain $m^2(0)\ll M^2$, two things
must happen:
\begin{itemize}
\item $m^2(M)\sim M^2$ - this is perfectly natural
\item $\frac{3}{4\pi^2}\lambda M^2$ must cancel $m^2(M)$ to high
accuracy.  There is no principle to enforce this.
\end{itemize}
This is the famous ``hierarchy problem.''

A supersymmetric theory has no quadratic divergences, so it solves
this ``technical'' aspect of the hierarchy problem.  Instead, one
is left with the ``aesthetic'' problem we already encountered
above. Fortunately, there are dynamical ways to solve this, so
this is a much softer version of the hierarchy problem.

To summarize, we have argued that the standard model successfully
explains why fermion and gauge boson masses are of ${\cal O}(v)$,
and hence much lighter than the hypothesized scale of fundamental
physics, $M$.  However, the standard model fails to explain why
the Higgs mass, or equivalently $v$, is much lighter than $M$ in
the first place. This ``hierarchy problem'' has occupied theorists
for a long time, and there are many potential solutions,
including:
\begin{itemize}
\item Perhaps $M$ is actually not much larger than $v$, so there
is no hierarchy after all.  Theories with extra dimensions are a
recent attempt along this direction.
\item Perhaps $M$ is only a little bigger than $v$, so it's really
only a ``little hierarchy.''  This is the rationale behind
``little Higgs'' models \cite{Schmaltz:2002wx}.
\item Perhaps $M$ really is much larger than $v$, and low-energy
supersymmetry makes this natural \cite{Martin:1997ns}.
\item Perhaps there are no fundamental scalars in nature after
all, and the electroweak symmetry is broken some other way. For
example, Technicolor models break the electroweak symmetry via a
fermion-antifermion condensate \cite{Hill:2002ap}.
\end{itemize}

If there really are no light fundamental scalars in nature, we
have an explanation at the ready.  Perhaps the hierarchy problem
is not a problem at all, but just a successful explanation of why
there are no light fundamental scalars.

No amount of deliberation by theorists alone is going to decide
which, if any, of these solutions is correct. Experiment will
decide. The Fermilab Tevatron is already probing the physics of
electroweak symmetry breaking, and the CERN Large Hadron Collider
is sure to give us a lot of information.  Our goal is to unravel
the mechanism of electroweak symmetry breaking, perhaps with the
assistance of a Linear ($e^+e^-$) Collider. It is going to take a
lot of effort on the part of both theorists and experimentalists,
and it is going to be an awful lot of fun.

\section*{Acknowledgments}

I would like to thank John Terning, Carlos Wagner, and Dieter
Zeppenfeld for the opportunity to present these lectures, and the
TASI students for their questions and comments. I am grateful for
assistance from A.~de~Gouvea, B.~Kayser, R.~Leigh, T.~McElmurry,
J.~Sayre, and S\"oren Wiesenfeldt. This work was supported in part
by the U.~S.~Department of Energy under contract
No.~DE-FG02-91ER40677.

\section*{Solutions to the exercises}

\subsection*{Section \ref{sec:weyl}}

\indent\indent {\em Exercise 1.1} -- For example,
\begin{eqnarray*}
[A_i,A_j]&=&\frac{1}{4}([J_i,J_j]+i[J_i,K_j]+i[K_i,J_j]-[K_i,K_j])\\
&=&\frac{1}{4}i\epsilon_{ijk}(J_k+iK_k+iK_k+J_k)\\
&=&i\epsilon_{ijk}A_k
\end{eqnarray*}
and similarly for the other commutators.
\\[7pt]
\indent {\em Exercise 1.2} -- The $(0,1/2)$ representation is
given by
\begin{eqnarray*}
A_i&=&0 \\
B_i&=&{1\over 2}\sigma_i
\end{eqnarray*}
which corresponds to
\begin{eqnarray*}
J_i&=&{1\over 2}\sigma_i \\
iK_i&=&-{1\over 2}\sigma_i\;.
\end{eqnarray*}
Hence the $(0,1/2)$ representation transforms under rotations and
boosts as
\begin{eqnarray*}
\chi &\to &e^{-{i\over 2} \sigma\cdot\theta}\chi \\
\chi &\to &e^{{1\over 2} \sigma\cdot\eta}\chi\;.
\end{eqnarray*}
Only the boost transformation differs from that of the $(1/2,0)$
representation, Eq.~(\ref{eq:rotboost}).
\\[7pt]
\indent {\em Exercise 1.3} -- Gluinos are the superpartners of
gluons, and therefore transform under the adjoint representation,
which is real.
\\[7pt]
\indent {\em Exercise 1.4} -- Under an $SU(2)$ transformation,
\begin{equation}
\epsilon_{ab}\chi^{aT}\epsilon\chi^{b} \to
\epsilon_{ab}U^a_c\chi^{cT}\epsilon U^b_d\chi^{d}\to
\epsilon_{cd}\chi^{cT}\epsilon\chi^{d}
\end{equation}
where I've used $\epsilon_{ab}U^a_cU^b_d=\det
U\epsilon_{cd}=\epsilon_{cd}$.  However,
\begin{equation}
(\chi^{aT}\epsilon\chi^{b})^T=\chi^{bT}\epsilon\chi^{a}
\end{equation}
where I've used the fact that the spinor fields are Grassmann
variables.  Hence $\chi^{aT}\epsilon\chi^{b}$ is symmetric under
interchange of $a,b$, and vanishes when contracted with
$\epsilon_{ab}$.

This exercise demonstrates a more general result: a Majorana mass
term cannot be constructed for a Weyl spinor that transforms under
a pseudoreal representation of a global or local symmetry, because
the tensor used to construct the mass term (the analogue of
$\epsilon_{ab}$) is necessarily antisymmetric
\cite{Peskin:1995ev}. A pseudoreal representation is one that is
unitarily equivalent to its complex conjugate: $U^*=SUS^\dagger$
for some unitary transformation $S$. For $SU(2)$, $S=\epsilon$.
\\[7pt]
\indent{\em Exercise 1.5} -- Under rotations,
\begin{equation}
\epsilon \xi^*  \to \epsilon M^*\xi^* =M\epsilon\xi^*
\end{equation}
where I have used the fact that $M=e^{-{i\over
2}\sigma\cdot\theta}$ is unitary if it is a rotation.  Under
boosts,
\begin{equation}
\epsilon \xi^*  \to \epsilon M^*\xi^* =M^{-1}\epsilon\xi^*
\end{equation}
where I have used the fact that $M=e^{-{1\over 2}\sigma\cdot\eta}$
is Hermitian if it is a boost.  This shows that $\epsilon\xi^*$
transforms under the $(0,1/2)$ representation of the Lorentz group
(see Exercise 1.2).
\\[7pt]
\indent{\em Exercise 1.6} -- Eq.~(\ref{eq:LMajorana}) may be
rewritten
\begin{eqnarray*}
-{1\over 2}m\bar\psi_M\psi_M&=&
-{1\over 2}m\left(\chi^\dagger,-\chi^T\epsilon\right)\left(\begin{array}{cc} 0 & 1 \\
1 & 0 \end{array}\right)\left(\begin{array}{c}\chi\\
\epsilon\chi^*\end{array}\right)\\
&=&{1\over 2}m(\chi^T\epsilon\chi-\chi^\dagger\epsilon\chi^*)
\end{eqnarray*}
which is equivalent to Eq.~(\ref{eq:Maj}).
\\[7pt]
\indent{\em Exercise 1.7} -- Eq.~(\ref{eq:LDiracpsic}) may be
rewritten
\begin{eqnarray*}
-m((\psi^c)_L^TC\psi_L + h.c.)
&=&-m\left(\left(\xi^T,0\right)\left(\begin{array}{cc} -\epsilon & 0 \\
0 & \epsilon \end{array}\right)\left(\begin{array}{c}\chi\\
0\end{array}\right)+h.c.\right)\\
&=&m(\xi^T\epsilon\chi+h.c.)
\end{eqnarray*}
which is Eq.~(\ref{eq:Dirac}).
\\[7pt]
\indent{\em Exercise 1.8} -- Eq.~(\ref{eq:LMajoranaDirac}) may be
rewritten
\begin{eqnarray*}
-{1\over 2}m(\psi_L^TC\psi_L + h.c.)
&=&-{1\over 2}m\left(\left(\chi^T,0\right)\left(\begin{array}{cc} -\epsilon & 0 \\
0 & \epsilon \end{array}\right)\left(\begin{array}{c}\chi\\
0\end{array}\right)+h.c.\right)\\
&=&{1\over 2}m(\chi^T\epsilon\chi+h.c.)
\end{eqnarray*}
which is Eq.~(\ref{eq:Maj}).
\\[7pt]
\indent{\em Exercise 1.9} -- Using the definition of a conjugate
spinor, we find
\begin{eqnarray*}
\psi_M^c&\equiv &C\gamma^0\psi_M^* \\
&=&\left(\begin{array}{cc} -\epsilon & 0 \\
0 & \epsilon \end{array}\right)\left(\begin{array}{cc} 0 & 1 \\
1 & 0 \end{array}\right)\left(\begin{array}{c}\chi^*\\ \epsilon\chi\end{array}\right)\\
&=&\left(\begin{array}{c}\chi\\
\epsilon\chi^*\end{array}\right)=\psi_M\;.
\end{eqnarray*}
\\[7pt]
\indent{\em Exercise 1.10} -- Eq.~(\ref{eq:LMajoranapsic}) may be
rewritten
\begin{eqnarray}
-{1\over 2}m(\overline{(\psi^c)}_R\psi_L + h.c.)
&=&-{1\over 2}m\left(\left(0,-\chi^T\epsilon\right)\left(\begin{array}{cc} 0 & 1 \\
1 & 0 \end{array}\right)\left(\begin{array}{c}\chi\\
0\end{array}\right)+h.c.\right)\\
&=&{1\over 2}m(\chi^T\epsilon\chi+h.c.)
\end{eqnarray}
which is Eq.~(\ref{eq:Maj}).

\subsection*{Section \ref{sec:flavor}}

\indent\indent {\em Exercise 2.1} -- Under an $SU(2)_L$
transformation,
\begin{equation}
\epsilon\phi^*\to \epsilon U^*\phi^*=U\epsilon\phi^*
\end{equation}
using the relation $\epsilon U^*=U \epsilon$ derived in Exercise
1.5 (in that case $U=M$ was a spatial rotation, but the group is
still $SU(2)$).
\\[7pt]
\indent{\em Exercise 2.2} -- The lepton sector has one complex
Yukawa matrix $\Gamma_e^{ij}$, and two $U(3)$ symmetries of the
matter Lagrangian, minus lepton number which is a symmetry of the
full Lagrangian, and cannot be used to diagonalize the mass
matrix. Thus the number of parameters is
\begin{equation}
2\times 3\times 3 - (3\times 3\times 2-1) = 1\;.
\end{equation}
However, we know that $m_e,m_\mu,m_\tau$ are independent, so there
are actually 3 parameters.  Thus there must be three symmetries of
the full Lagrangian that we need to subtract from the $U(3)$
symmetries, not just one.  These are the individual lepton numbers
$L_e,L_\mu,L_\tau$ (where $L=L_e+L_\mu+L_\tau$).
\\[7pt]
\indent{\em Exercise 2.3} -- $L_L$ and $\phi$ have equal and
opposite hypercharge, so $N_R$ must have zero hypercharge in order
for ${\cal L}_{Yukawa}$ to be gauge invariant.  Similarly, $L_L$
and $\epsilon\phi^*$ both transform as $SU(2)_L$ doublets, so
${\cal L}_{Yukawa}$ is gauge invariant if $N_R$ is inert under
$SU(2)_L$.  In order for ${\cal L}_{Yukawa}$ to conserve lepton
number, we must assign $L=+1$ to $N_R$.
\\[7pt]
\indent{\em Exercise 2.4} -- The combination $L_L^T\epsilon\phi$
is $SU(2)_L\times U(1)_Y$ invariant (see Exercise 2.3), so ${\cal
L}_5$ is gauge invariant.  Lepton number is violated because $L_L$
carries $L=+1$, so ${\cal L}_5$ has $L=+2$.
\\[7pt]
\indent{\em Exercise 2.5} -- The counting is similar to that of
Exercise 2.2, with two differences.  First, we must add the
complex, symmetric matrix $c^{ij}$, which has $2\times 6$
parameters. Second, we should not subtract lepton number from the
symmetries of the matter Lagrangian, because it is no longer a
symmetry of the full Lagrangian, being violated by ${\cal L}_5$.
Hence the number of physically relevant parameters is
\begin{equation}
2\times 3\times 3 + 2\times 6 - 3\times 3\times 2 = 12\;.
\end{equation}
Of these parameters, six are the charged-lepton and neutrino
masses, leaving six parameters for the MNS matrix. Three are
mixing angles, and three are $CP$-violating phases.
\\[7pt]
\indent{\em Exercise 2.6} -- It's easiest to do this with
index-free notation:
\begin{equation}
\frac{\partial{\cal L}}{\partial N_R}=-\bar L_L\Gamma_\nu\epsilon
\phi^* -N_R^{T}M_RC+h.c.\;.
\end{equation}
Solving for $N_R$ gives
\begin{equation}
N_R=\phi^\dagger\epsilon C\gamma^0(\Gamma_\nu M_R^{-1})^TL_L^*\;.
\end{equation}
Plugging this back into ${\cal L}$ to eliminate $N_R$ gives
\begin{equation}
{\cal L}={1\over 2}L_L^\dagger\epsilon\phi^*C\Gamma_\nu(\Gamma_\nu
M_R^{-1})^T\phi^\dagger\epsilon L^* + h.c.\;.
\end{equation}
This is equal to Eq.~(\ref{eq:L5}), where we identify the first
term above with the $h.c.$ term of ${\cal L}_5$.  This gives
\begin{equation}
\frac{c^\dagger}{M}=-{1\over 2}\Gamma_\nu(\Gamma_\nu M_R^{-1})^T
\end{equation}
which is equivalent to Eq.~(\ref{eq:cM}) if we recall that $c$ is
a symmetric matrix.

\subsection*{Section \ref{sec:custodial}}

\indent\indent {\em Exercise 3.1} -- Writing a matrix as an
outer-product of two vectors,
\begin{equation}
{\rm Tr}\;\Phi^\dagger\Phi=\frac{1}{2}{\rm
Tr}\;\left(\begin{array}{c}-\phi^T\epsilon\\
\phi^\dagger\end{array}\right)(\epsilon\phi^*,\phi) =
\phi^\dagger\phi
\end{equation}
where I've used $\phi^T\phi^*=(\phi^T\phi^*)^T=\phi^\dagger\phi$.
This shows that the potentials are equivalent. For the
gauge-covariant kinetic-energy term, we need
\begin{equation}
D_\mu\Phi =
(D_\mu(\epsilon\phi^*),D_\mu\phi)=(\epsilon(D_\mu\phi)^*,D_\mu\phi)
\end{equation}
where the last step uses $\sigma\epsilon = -\epsilon\sigma^T
=-\epsilon\sigma^*$.  From there the proof proceeds as above.
\\[7pt]
\indent{\em Exercise 3.2} -- Consider the $U(1)$ subgroup of
$SU(2)_R$ given by $e^{{i\over 2}\sigma_3\theta}$.  Inserting this
for the transformation $R$ in Eq.~(\ref{eq:SU2R}) yields the
hypercharge transformation of Eq.~(\ref{eq:gauge}).
\\[7pt]
\indent{\em Exercise 3.3} -- $W_\mu^A$ transforms under an
$SU(2)_L$ gauge transformation as
\begin{equation}
{\sigma\over 2}\cdot W_\mu\to L{\sigma\over 2}\cdot W_\mu
L^\dagger + \frac{i}{g}(\partial_\mu L)L^\dagger\;.
\end{equation}
Under a global $SU(2)_L$ transformation, the second term is
absent, and $W_\mu^A$ transforms simply under the adjoint
(triplet) representation of $SU(2)_L$.  Since $W_\mu^A$ is a
singlet under $SU(2)_R$, it also transforms as a triplet under
$SU(2)_{L+R}$.
\\[7pt]
\indent{\em Exercise 3.4} -- Using l'H\^{o}pital's rule for sick
functions gives
\begin{equation}
\lim_{x\to 1} \frac{\ln x}{x-1}=\lim_{x\to 1} {1\over x}=1
\end{equation}
where $x=m_t^2/m_b^2$.  Thus the correction, which is proportional
to
\begin{equation}
x+1-2\frac{x}{x-1}\ln x
\end{equation}
vanishes in the limit $x\to 1$.
\\[7pt]
\indent{\em Exercise 3.5} -- In the limit $m_t=m_b$, the top and
bottom Yukawa couplings are equal, and the Yukawa Lagrangian of
Eq.~(\ref{eq:LYukawa}) (restricted to the third generation of
quarks) may be written
\begin{equation}
{\cal L}_{Yukawa}= -y\bar Q_L\Phi Q_R+h.c.
\end{equation}
where $Q_R=(t_R, b_R)$ is the analogue of $Q_L$ for the
right-chiral quark fields.  Under $SU(2)_L$, $Q_L \to LQ_L$. If we
let $Q_R$ transform under $SU(2)_R$ as $Q_R\to RQ_R$, then the
Yukawa Lagrangian is invariant under $SU(2)_L\times SU(2)_R$
custodial symmetry:
\begin{equation}
\bar Q_L\Phi Q_R \to \bar Q_LL^\dagger L\Phi R^\dagger RQ_R = \bar
Q_L\Phi Q_R\;.
\end{equation}
\indent{\em Exercise 3.6} -- Under an infinitesimal $SU(2)_L$
transformation,
\begin{eqnarray*}
e^{i\frac{\sigma\cdot\pi}{v}}&\to &e^{{i\over
2}\sigma\cdot\theta}e^{i\frac{\sigma\cdot\pi}{v}}\\
1+i\frac{\sigma\cdot\pi}{v}+\cdots &\to &\left(1+{i\over
2}\sigma\cdot\theta+\cdots\right)\left(1+i\frac{\sigma\cdot\pi}{v}+\cdots\right)\\
\pi_i &\to &\frac{v}{2}\theta_i + \pi_i
\end{eqnarray*}
while under an infinitesimal $SU(2)_{L+R}$ transformation
\begin{eqnarray*}
e^{i\frac{\sigma\cdot\pi}{v}}&\to &e^{{i\over
2}\sigma\cdot\theta}e^{i\frac{\sigma\cdot\pi}{v}}e^{-{i\over
2}\sigma\cdot\theta}\\
1+i\frac{\sigma\cdot\pi}{v}+\cdots &\to &\left(1+{i\over
2}\sigma\cdot\theta+\cdots\right)\left(1+i\frac{\sigma\cdot\pi}{v}+\cdots\right)\left(1-{i\over
2}\sigma\cdot\theta+\cdots\right)\\
&=&1+i\frac{\sigma\cdot\pi}{v}-{1\over 2v}\theta_i\pi_j[\sigma_i,\sigma_j]+\cdots\\
\pi_i&\to &\pi_i-\epsilon_{ijk}\theta_j\pi_k
\end{eqnarray*}
where I've used $[\sigma_i,\sigma_j]=2i\epsilon_{ijk}\sigma_k$.
\\[7pt]
\indent{\em Exercise 3.7} -- The operator is invariant under
global $SU(2)_L$ (as it must be, since it is also invariant under
local $SU(2)_L$),
\begin{equation}
{\rm Tr}\;\sigma_3\Sigma^\dagger D_\mu\Sigma \to {\rm
Tr}\;\sigma_3\Sigma^\dagger L^\dagger D_\mu L\Sigma = {\rm
Tr}\;\sigma_3\Sigma^\dagger D_\mu\Sigma
\end{equation}
but not under $SU(2)_R$,
\begin{equation}
{\rm Tr}\;\sigma_3\Sigma^\dagger D_\mu\Sigma \to {\rm
Tr}\;\sigma_3R\Sigma^\dagger D_\mu \Sigma R^\dagger \neq {\rm
Tr}\;\sigma_3\Sigma^\dagger D_\mu\Sigma
\end{equation}
because of the presence of $\sigma_3$.  Only the hypercharge
subgroup of $SU(2)_R$ is respected (see Exercise 3.2), as it must
be.



\newpage

\begin{thebibliography}{99}

\bibitem{Peskin:1995ev}
M.~E.~Peskin and D.~V.~Schroeder, {\sl An Introduction to Quantum
Field Theory} (Addison-Wesley, Reading, 1995).

\bibitem{Weinberg:1995mt}
S.~Weinberg, {\sl The Quantum Theory of Fields, Vol. 1:
Foundations} (Cambridge, 1995).

\bibitem{Wess:1992cp}
J.~Wess and J.~Bagger, {\sl Supersymmetry and Supergravity}
(Princeton, 1992).

\bibitem{Kayser:1989iu}
B.~Kayser, F.~Gibrat-Debu and F.~Perrier, {\sl The Physics Of
Massive Neutrinos}, World Sci.\ Lect.\ Notes Phys.\ {\bf 25}, 1
(1989).

\bibitem{Falk:2000tx}
A.~F.~Falk, ``The CKM matrix and the heavy quark expansion,'' in
{\sl Flavor Physics for the Millennium}, TASI 2000, ed.~J.~Rosner
(World Scientific, Singapore, 2001), p.~379
[arXiv:hep-ph/0007339].

\bibitem{Georgi:1994qn}
H.~Georgi, ``Effective field theory,'' Ann.\ Rev.\ Nucl.\ Part.\
Sci.\  {\bf 43}, 209 (1993).

\bibitem{Sikivie:1980hm}
P.~Sikivie, L.~Susskind, M.~B.~Voloshin and V.~I.~Zakharov,
``Isospin Breaking In Technicolor Models,'' Nucl.\ Phys.\ B {\bf
173}, 189 (1980).

\bibitem{Veltman:1994wz}
M.~J.~G.~Veltman, {\sl Diagrammatica: The Path to Feynman Rules}
(Cambridge, 1994).

\bibitem{Chivukula:1995hr}
R.~S.~Chivukula, M.~J.~Dugan, M.~Golden and E.~H.~Simmons,
``Theory of a strongly interacting electroweak symmetry breaking
sector,'' Ann.\ Rev.\ Nucl.\ Part.\ Sci.\  {\bf 45}, 255 (1995)
[arXiv:hep-ph/9503230].

\bibitem{Eidelman:2004wy}
S.~Eidelman {\it et al.}  [Particle Data Group Collaboration],
Phys.\ Lett.\ B {\bf 592}, 1 (2004).

\bibitem{Schwarz:1998ny}
J.~H.~Schwarz and N.~Seiberg, ``String theory, supersymmetry,
unification, and all that,'' Rev.\ Mod.\ Phys.\  {\bf 71}, S112
(1999) [arXiv:hep-th/9803179].

\bibitem{Schmaltz:2002wx}
M.~Schmaltz, ``Physics beyond the standard model (Theory):
Introducing the little Higgs,'' Nucl.\ Phys.\ Proc.\ Suppl.\  {\bf
117}, 40 (2003) [arXiv:hep-ph/0210415].

\bibitem{Martin:1997ns}
S.~P.~Martin, ``A supersymmetry primer,'' in {\sl Perspectives on
Supersymmetry}, ed.~G.~Kane (World Scientific, Singapore, 1998),
p.~1 [arXiv:hep-ph/9709356].

\bibitem{Hill:2002ap}
C.~T.~Hill and E.~H.~Simmons, ``Strong dynamics and electroweak
symmetry breaking,'' Phys.\ Rept.\  {\bf 381}, 235 (2003)
[Erratum-ibid.\  {\bf 390}, 553 (2004)] [arXiv:hep-ph/0203079].

\end{thebibliography}
\end{document}